\documentstyle[12pt,titlepage]{article}
\begin{document}
\pagestyle{empty}
\newcommand{\lapproxeq}{\lower
.7ex\hbox{$\;\stackrel{\textstyle <}{\sim}\;$}}
\baselineskip=0.212in

\begin{flushleft}
\large
{MKPH-T-92-18  \hfill November 10, 1992}  \\
\end{flushleft}

\vspace{1.8cm}

\begin{center}

\Large{{\bf Test of Scalar Meson Structure
            in $\bf \phi$ Radiative Decays}} \\

\vspace{1.8cm}

\Large
{S. Kumano $^\star$ }         \\

\vspace{1.3cm}

\Large
{Institut f\"ur Kernphysik, Universit\"at Mainz}    \\

\vspace{0.1cm}

\Large
{6500 Mainz, Germany $^{\star\star}$}         \\

\vspace{0.1cm}

\Large
{and}      \\

\vspace{0.1cm}

\Large
{Institute for Nuclear Study, University of Tokyo} \\

\vspace{0.1cm}

\Large
{Midori-cho, Tanashi, Tokyo 188, Japan} \\

\vspace{2.5cm}

\large
{Talk given at the International Workshop on Flavor and Spin } \\

{in Hadronic and Electromagnetic Interactions} \\

{Torino, Italy, September 21-23, 1992}  \\

\end{center}

\vspace{2.7cm}

{\rule{6.cm}{0.1mm}} \\

\vspace{-0.4cm}

\normalsize
{$\star$ research supported in part by
         the Deutsche Forschungsgemeinschaft (SFB 201).} \\

\vspace{-0.4cm}

\normalsize
{$\star\star$ present address. E-mail:
              KUMANO@VKPMZP.KPH.UNI$-$MAINZ.DE}  \\

\vfill\eject
\pagestyle{plain}
\begin{center}

\Large
{TEST OF SCALAR MESON STRUCTURE} \\

\Large
{IN $\phi$ RADIATIVE DECAYS}      \\

\vspace{0.5cm}

{S. Kumano}             \\

{Institut f\"ur Kernphysik, Universit\"at Mainz}    \\

{6500 Mainz, Germany $^*$}      \\

{and}

{Institute for Nuclear Study, University of Tokyo} \\

{Midori-cho, Tanashi, Tokyo 188, Japan}

\vspace{0.7cm}

\normalsize
Abstract
\end{center}
\vspace{-0.46cm}

We show that $\phi$ radiative decays
into scalar mesons
[$f_0$(975),$a_0$(980)$\equiv S$] can
provide important clues on
the internal structures of these mesons.
Radiative decay widths
vary widely: $B.R.=10^{-4}-10^{-6}$
depending on the substructures
($q\bar q$, $qq\bar q\bar q$, $K\bar K$, glueball).
Hence, we could discriminate among various models
by measuring these widths at future $\phi$ factories.
The understanding of these meson structures is valuable
not only in hadron spectroscopy but also
in nuclear physics in connection with the widely-used
but little-understood $\sigma$ meson.
We also find that the decay
$\phi\rightarrow S\gamma \rightarrow K^0 \bar K^0\gamma$
is not strong
enough to pose a significant background problem
for studying CP violation
via $\phi\rightarrow K^0 \bar K^0$
at the $\phi$ factories.

\vspace{0.8cm}

\noindent
{1. \underline{Introduction to $\phi \rightarrow S \gamma$}}

\vspace{0.2cm}

This talk is based on the research done with
F. E. Close and N. Isgur.
For the details of our results presented
in this paper, readers are suggested
to read the joint paper in Ref. 1.

There are two major purposes for studying
$\phi$ radiative decays. One is to understand
scalar-meson [$f_0(975)$ and $a_0(980)$]
structures and the other is to investigate
a possible background problem for
studies of CP violation at
future $\phi$ factories.

Meson spectroscopy in the 1 GeV region
has been well investigated and most mesons
can be explained by naive $q\bar q$ models [2].
However, structures of the scalar mesons $f_0$
and $a_0$ (which we denote as $S$ in this paper)
are not well understood.
There are a few ``evidences'' against identifying
them with the $q\bar q$-type mesons.
For example, if the strong-decay width of $f_0$
is calculated by assuming the $q\bar q$ structure,
we obtain 500$-$1000 MeV width [3], which is in
contradiction to the experimental one, 33.6 MeV [4].
Even if ambiguity of the hadronic matrix element
is taken into account, the difference of an order
of magnitude is too large.
On the other hand, the scalar meson $\sigma$ with
the width of 500$-$1000 MeV has been widely used
in nuclear physics [5]. Its large width is essential
for explaining isoscalar $\pi\pi$ phase shifts [6].
There are other evidences against the
$q \bar q$ picture, for example,
in studies of $\gamma \gamma$ couplings of $S$ [7,8].
Considering these circumstances, we guess
that the ordinary $^3P_0$ $q\bar q$ meson corresponds
to the $\sigma$ meson and the observed $f_0(975)$
is a more exotic meson, such as
$qq\bar q\bar q$, $K\bar K$, or a glueball.
The situation is also similar in
the isovector-partner $a_0$-meson case [1].
It is important to understand the structures of
these mesons not only in
hadron spectroscopy but also
in nuclear physics
in connection  with the $\sigma$ meson.

There are $\phi$ factory proposals [9]
at Frascati, KEK, Novosibirsk, and UCLA.
We expect that some $\phi$ factories
will be built in several years.
One of our research purposes is to show that
the scalar-meson structures could be
investigated at the $\phi$ factories.
Because the spin and parity of $\phi$ are $1^-$
and those of $S$ are $0^+$, the radiative decay
$\phi\rightarrow S\gamma$ is an electric-dipole (E1)
decay. The electric-dipole operator is given by
$\sum e_i \vec r_i$, where $e_i$ is the
charge of a constituent and $\vec r_i$ is
the vector distance from the center of mass.
Therefore, its matrix element is very sensitive
to the scalar-meson size. If $S$ were the $q \bar q$ meson,
charges should be confined in rather small space
(about 0.5 fm size).
On the other hand, if $S$ is
the $K\bar K$ molecule [6], the charges should be
distributed in larger space
(about 1.5 fm size).
This is because scalar-meson masses are
just below the $K\bar K$ threshold
(2$m_{_{K^\pm}}$=987 MeV)
and binding energies are small.
There is an analogous case in nuclear physics.
The deuteron exists slightly below the p-n
threshold; hence it is a very loose bound state.
{}From the above discussions, we can reasonably
expect that charge distributions in the $q\bar q$ and
$K\bar K$ cases are much different. The E1-decay width
of $\phi\rightarrow S\gamma$ shall clearly reflect
the difference. In this paper, we show our
decay-width calculations based on the various models
for the scalar mesons.

One of the major purposes for building the $\phi$ factories
is to measure CP violation parameters accurately
by using the decay $\phi\rightarrow K_S K_L$[10].
CP violation effects are small, so
we should be careful in excluding possible backgrounds.
An important background due to the
radiative decay
$\phi\rightarrow S\gamma\rightarrow K^0 \bar K^0 \gamma$
was recently pointed out [11,1,12].
Because the photon charge conjugation is negative,
we have $K_S K_S$ or $K_L K_L$ instead of $K_S K_L$.
If the branching ratio for the radiative decay is large
[B.R.$(\phi\rightarrow K^0 \bar K^0 \gamma) >10^{-6}$],
it is a significant background
to the studies of CP violation.
At this stage, there are several theoretical
predictions for this branching ratio; however,
they vary several orders of
magnitude (B.R.=$10^{-5}-10^{-9}$) [11].
It is important to reach a theoretical agreement
on the magnitude before we start
the CP-violation experiments
at the $\phi$ factories.

In section 2, the decay widths for $\phi\rightarrow S\gamma$
are estimated by using various models for the scalar mesons.
In  section 3, our investigation of the decay
$\phi\rightarrow S\gamma$ through a $K\bar K$
loop is discussed by focusing on
the $K\bar K$-molecule picture for $S$.
Comments on the CP-background problem
and on the OZI rule are given in section 4, and
conclusions are in section 5.

\vfill\eject

\noindent
{2. \underline{Estimates of decay widths
    $\Gamma (\phi \rightarrow S\gamma )$
    in various models for $S$}}

\vspace{0.2cm}

Assuming various models for $S$
($n \bar n$, $s \bar s$, $qq\bar q\bar q$,
 $K\bar K$, and a glueball),
where $n$=$u$ or $d$-quark, we calculate
widths for the $\phi$ radiative decays.
It should be noted that this type of investigation
is still at the early stage, so that
our calculations are naive
order-of-magnitude estimates.
Our purpose is simply to show that the decay widths
are very dependent on the $S$ structure; hence
we could determine the structure by measuring
them at the $\phi$ factories.

$~~~~~$

\noindent
$\bullet$ $f_0=s\bar s$ and $a_0=(u\bar u-d\bar d)/\sqrt 2$

If $f_0$ is a $s\bar s$ bound state and
   $a_0$ is an ordinary $n\bar n$ state, we first estimate
the decay width of $\phi \rightarrow f_0(s\bar s)\gamma$
by using observed E1-decay widths.
For example, the charmonium E1 decay
$\chi_{c0}\rightarrow J/\psi \gamma$
is observed and the experimental width is 0.092 MeV.
E1 decay widths are in general expressed as
$\Gamma (E1) \sim e^2 R^2 E_\gamma^3$, where
$e$ is the charge factor, $R$ is the hadron size, and
$E_\gamma$ is the emitted-photon energy.
Taking into account these factors, we obtain
$$
B.R.(\phi\rightarrow f_0\gamma)\approx
{1 \over {4.41}} \cdot {1 \over 3} \cdot
\bigl ( {{e_s} \over {e_c}} \bigr )^2 \cdot
\bigl ( {{R_s} \over {R_c}} \bigr )^2 \cdot
\bigl ( {{44}  \over {320}} \bigr )^3 \cdot
0.092 \simeq 1 \times 10^{-5} ~~~,
\eqno{(1.1)}
$$
where 1/3 is a $\phi$-spin factor and 4.41 MeV is
the total $\phi$ decay width.
We also used other E1 decays
(e.g. $b_1(1235)\rightarrow\pi\gamma$)
and obtained similar numerical results.
The radiative decay into $a_0$ is an OZI violating
process, so that the branching ratio should be significantly
smaller than the one in Eq. (1.1):
$$
R(a_0/f_0)\equiv \Gamma (\phi \rightarrow a_0\gamma)/
                  \Gamma (\phi \rightarrow f_0\gamma)
 << 1 ~~~.
\eqno{(1.2)}
$$

$~~~~~$

\noindent
$\bullet$ $f_0=(u\bar u+d\bar d)/\sqrt 2$ and
          $a_0=(u\bar u-d\bar d)/\sqrt 2$

If both $f_0$ and $a_0$ are ordinary $n\bar n$ type mesons,
both radiative decays are OZI-violating processes.
Therefore, the decay widths are much smaller than
the value in Eq. (1.1):
$$
\Gamma (\phi\rightarrow f_0\gamma),
\Gamma (\phi\rightarrow a_0\gamma) \lapproxeq 10^{-6} ~~~.
\eqno{(2)}
$$
Absolute values should be calculated
by considering the $\phi-\omega$ mixing.

$~~~~~$

\noindent
$\bullet$ $qq\bar q\bar q$

If the scalar mesons are four-quark bound states
($qq\bar q\bar q$) [13], we find that the decay-width
ratio $R(a_0/f_0)$ provides us important information
on the internal structure.
The decay widths themselves are rather difficult to
be estimated due to poorly-understood
decays of four-quark bound states.
In the following, we show that
the ratio $R(a_0/f_0)$ varies
significantly with
the $qq\bar q\bar q$ structure.
Because the electric-dipole operator
is given by $\sum e_i \vec r_i$,
we simply take into account constituent charges
for evaluating the widths.

\noindent
$\bullet\bullet$ $qq\bar q\bar q$
                 [$K\bar K$-like ``bag''=$(n\bar s)(\bar n s)$]

If the $f_0$ and $a_0$ structures are
$K \bar K$-like ``bag'', we denote
$S=(u\bar s)(\bar u s)\pm
           (d\bar s)(\bar d s)$,
where $+$ is for $f_0$ and $-$ for $a_0$.
In this case, E1 matrix elements are roughly given by
$M(E1)\sim (e_u+e_{\bar s})\pm
           (e_d+e_{\bar s})=1$.
Because the second term vanishes ($e_d+e_{\bar s}=0$),
we obtain the same matrix elements for both
decays. So, the decay-width ratio is
$$
R(a_0/f_0)\approx 1 ~~~.
\eqno{(3)}
$$

\noindent
$\bullet\bullet$ $qq\bar q\bar q$
                 [$D\bar D$-like ``bag''=$(ns)(\bar n \bar s)$]

If the structures are diquark-antidiquark-like
``bag'', we denote
        $S=(us)(\bar u \bar s)\pm
           (ds)(\bar d \bar s)$.
The E1 matrix elements are given by
$M(E1)\sim (e_u+e_s)\pm
           (e_d+e_s)=-1/3 ~(f_0),~1~(a_0)$ and
the ratio is
$$
R(a_0/f_0)\approx 9 ~~~,
\eqno{(4)}
$$
which is
much different from the one
in the $K\bar K$-like bag case.
Therefore, we find that the decay-width ratio
is also an important quantity for investigating
the structures.

\noindent
$\bullet\bullet$ $qq\bar q\bar q$
                 [$\pi\eta$, $\eta\eta$-like ``bag''=
                  $(n\bar n)(s\bar s)$]

If the structures are
$\pi\eta$ or $\eta\eta$-like ``bag'',
we denote $S=(u\bar u)(s\bar s)\pm
           (d\bar d)(s\bar s)$.
The E1 matrix elements are given by
$M(E1)\sim [(e_u+e_{\bar u})-(e_s+e_{\bar s})]
       \pm [(e_d+e_{\bar d})-(e_s+e_{\bar s})]=0$,
so that we cannot obtain the ratio by the above simple
method. Much more detailed analyses are needed for
the decay-width ratio and the widths themselves.

$~~~~~$

\noindent
$\bullet$ $K\bar K$ molecule (``diffuse $K \bar K$'')

The constituent-quark contents of
the $qq\bar q\bar q$ and $K\bar K$ systems
are identical, but their dynamical
structures are very different. For example, it is clear that
the deuteron should be regarded as the $p$-$n$ bound state
but not as a six-quark-bag-like bound state.
The essential feature is whether the multiquark
system is confined within a hadronic system
with a radius of the order of 1/$\Lambda_{QCD}$ or it is
two identifiable color singlets spread over
a region significantly greater than this.
Decay-width calculations based on a $K\bar K$-molecule
picture are discussed in section 3. The results indicate
$$
B.R.(\phi\rightarrow f_0\gamma)\simeq 4 \times 10^{-5}
{}~~~,~~~
R(a_0/f_0)\approx 1
{}~~~.
\eqno{(5)}
$$
This rate confirms our original expectation
in the introduction that the E1-decay width should
reflect the size of $S$; therefore,
the width for the $K\bar K$ is larger than
the ones for the $q\bar q$ models.

\vfill\eject

\noindent
$\bullet$ $f_0$=glueball

If $f_0$ is a glueball, we estimate the quarkonium-glueball
mixing by the observed $f_0\rightarrow\pi\pi$ decay width and
the calculated $^3P_0(q\bar q)\rightarrow\pi\pi$ width [3]:
the mixing$\lapproxeq \Gamma (f_0\rightarrow\pi\pi)/
                      \Gamma (^3P_0(q\bar q)\rightarrow\pi\pi)
           =26/(500-1000)
           \lapproxeq 1/20$.
Therefore, the decay width is
      $\Gamma(\phi\rightarrow f_0(glueball)\gamma)
           \lapproxeq
       \Gamma(\phi\rightarrow f_0(q\bar q)\gamma)/20$.
Using the results in Eqs. (1.1) and (2)
for $\Gamma(\phi\rightarrow f_0(q\bar q)\gamma)$,
we obtain
$$
B.R.(\phi \rightarrow f_0\gamma)
           \lapproxeq 10^{-6} ~~~.
\eqno{(6)}
$$

$~~~~~$

As we found in this section,
not only the decay width $\Gamma(\phi\rightarrow f_o\gamma)$
but also the ratio $\Gamma(\phi\rightarrow a_0\gamma)/
                    \Gamma(\phi\rightarrow f_0\gamma)$
is important for discriminating among the models.
The obtained results are summarized in the following
table.

\vspace{1.0cm}
\begin{center}
\normalsize
\begin{tabular}{|cc|c|l|}   \hline
{}~~~            &  ~~~  & Ratio & ~~~ \\
\multicolumn{2}{|c|}{Constitution}
                & $\Gamma (\phi\rightarrow\gamma a_0)/~$
                & \multicolumn{1}{|c|}{Absolute B.R.}            \\
{}~~~ & ~~~ & $~\Gamma (\phi\rightarrow\gamma f_0)$ & ~~~   \\ \hline
{}~~~ & ~~~ & ~~~ & ~~~ \\
$K\bar K$      & molecule=``diffuse $K\bar K$''
               & 1
               & $a_0\simeq f_0\simeq 4\times 10^{-5}$       \\
{}~~~ & ~~~ & ~~~ & ~~~ \\ \hline
{}~~~ & ~~~ & ~~~ & ~~~ \\
$q^2\bar q^2$  & $K\bar K$-like ``bag''$=(n\bar s)(\bar n s)$
               & 1
               & ~~~  \\
{}~~~ & ~~~ & ~~~ & ~~~ \\ \cline{2-3}
{}~~~ & ~~~ & ~~~ & $a_0,f_0\lapproxeq 10^{-6}$ ? \\
{}~~~            & $D\bar D$-like ``bag''$=(ns)(\bar n \bar s)$
               & 9
               & ~~~ \\
{}~~~ & ~~~ & ~~~ & ~~~ \\ \cline{2-3}
{}~~~ & ~~~ & ~~~ & ~~~~see Ref. 1 \\
{}~~~            & $\pi\pi,\pi\eta$-like ``bag''$=(n\bar n)(s\bar s)$
               & $-$
               & ~~~~~ \\
{}~~~ & ~~~ & ~~~ & ~~~ \\ \hline
{}~~~ & ~~~ & ~~~ & ~~~ \\
$^3P_0(q\bar q)$  & $f_0(n\bar n)$, $a_0(n\bar n)$
                  & $-$
                  & $a_0,f_0\lapproxeq 10^{-6}$   \\
{}~~~ & ~~~ & ~~~ & ~~~ \\ \cline{2-4}
{}~~~ & ~~~ & ~~~ & ~~~ \\
{}~~~               & $f_0(s\bar s)$, $a_0(n\bar n)$
                  & $\approx$ 0
                  & $f_0 \simeq 1 \times 10^{-5}$        \\
{}~~~ & ~~~ & ~~~ & ~~~ \\ \hline
{}~~~ & ~~~ & ~~~ & ~~~ \\
glueball          & $f_0$
                  & $-$
                  & $f_0\lapproxeq 10^{-6}$     \\
{}~~~ & ~~~ & ~~~ & ~~~ \\ \hline
\end{tabular}

\vspace{1.0cm}

Table 1. Summary of possibilities
\end{center}

\vfill\eject

\noindent
{3. \underline{$\phi\rightarrow K\bar K \gamma \rightarrow S \gamma$
    in a $K\bar K$ model for $S$}}

\vspace{0.2cm}

We discuss our calculation of the decay width
$\Gamma (\phi\rightarrow S\gamma)$ based on
a $K\bar K$ model for $S$.
There are existing literatures on the decay
through a $K\bar K$ loop [11].
All of these calculations assume a pointlike
coupling in the $SK\bar K$ coupling.
However, it is not an appropriate
description especially if the scalar meson $S$
is the $K\bar K$ molecule.
Because it is a loose bound state of $K$ and $\bar K$,
the keon-loop momentum cannot be an infinite quantity.
The momentum should be restricted by the size of
the bound system. In order to take into account
of this effect, we introduce a momentum cutoff
given by a momentum-space wave function
for the $K\bar K$ molecule.
For simplicity in our numerical analysis, we take
the cutoff given by the dipole form factor:
$
\phi (|\vec k|)=1/(1+\vec k^2/\mu^2)^2,
$
where the cutoff parameter $\mu$ is given by
the $K\bar K$-molecule radius $R_{K\bar K}$ as
$\mu=\sqrt{3}/(2R_{K\bar K})$.
If we take the radius 1.2 fm, this cutoff
is essentially the $K\bar K$-molecule
wave function obtained by a Toronto group [7,6].

Once such momentum dependence is introduced,
we have to be careful in current conservation
or gauge invariance. By the minimal substitution
$\phi(\vec k) \rightarrow \phi(\vec k -e\vec A)$, we
obtain a new current, so called ``interaction current''.
The physics meaning of this current is
as follows [14].
The finite-range form factor $\phi(k)$
at the $SK\bar K$ vertex means that
$K$ and $\bar K$ are annihilated into
$S$ with a finite distance $r$, which
is controlled by the cutoff
$\psi (r)=\int d^3k e^{i\vec k\cdot \vec r} \phi(|\vec k|)$.
Therefore, current flows associated with this finite distance
during $K\bar K\rightarrow S$ must be included
in order to satisfy the current conservation.
This is the physics meaning of the new current
shown in Fig. 1d.
Fortunately, photon energies in the $\phi$ radiative decays
are about 40 MeV, which is considered to be a
very-long-wavelength region in hadronic scale.
Therefore, we simply expand
$H_{SK\bar K}=g\phi(\vec k-e\vec A)$ in
the Taylor series and take the second term for obtaining
the interaction current. Such a prescription
should be good enough due to the soft photon [14].
In this way, we obtain the interaction current as
$$
J_{int}^\mu=eg \phi'(|\vec k|)\hat k^\mu ~~~,
\eqno{(7)}
$$
where $\hat k=(0,\vec k/|\vec k|)$.

Now, we are in good preparation for calculating
the decay width.
Matrix elements in Fig. 1 are written as
$$
M_a^\mu ~=~ - e g g_\phi \int {{d^4 k} \over {(2\pi)^4}}
             ~ \phi (|\vec k|) ~
              {{2 \varepsilon _\phi^\mu} \over {D(k-q/2)D(k+q/2-p)}}
{}~~~~~~~~~~~~~~~~~~
\hfill
\eqno{(8.1)}
$$
$$
M_b^\mu ~=~ + e g g_\phi \int {{d^4 k} \over {(2\pi)^4}}
              ~\phi (|\vec k|) ~
              {{\varepsilon _\phi \cdot (2k+q-p) ~(2k)^\mu}
                \over {D(k+q/2)D(k-q/2)D(k+q/2-p)}}
{}~~~~
\eqno{(8.2)}
$$
$$
M_c^\mu ~=~ + e g g_\phi \int {{d^4 k} \over {(2\pi)^4}}
            ~  \phi (|\vec k|) ~
              {{\varepsilon _\phi \cdot (2k-q+p)~ (2k)^\mu}
                \over {D(k+q/2)D(k-q/2)D(k-q/2+p)}}
{}~~~~
\eqno{(8.3)}
$$
$$
M_d^\mu ~=~ + e g g_\phi \int {{d^4 k} \over {(2\pi)^4}}
             ~ \phi ' (|\vec k|) ~
              {{\varepsilon _\phi \cdot (2k-p) ~\hat k ^\mu}
                \over {D(k)D(k-p)}}
{}~~~~~~~~~~~~~~~~~~~~~~~~~~~~~~~~~~
\eqno{(8.4)}
$$
where $p$, $q$, and $k$ are the
$\phi$, photon, and keon-loop momenta respectively,
and $D(k)$ is defined by $D(k)~=~ k^2 - m_{_{K}} ^2$.
In the particular case where $\phi (|\vec k|)=1$ and
$\phi ' (|\vec k|)=0$, these reproduce the familiar
field-theory expressions of Ref. 11.
It is interesting to note the role that
$\phi ' (|\vec k|)$ plays in regularizing the
infinite integral.
Using these expressions, we can check explicitly
the current conservation $\sum q\cdot M_j$=0.
We define the matrix elements $\tilde M_j$ ($j=a-d$)
by $\tilde M_j = \varepsilon_\gamma \cdot M_j
 /(ie\varepsilon_\gamma \cdot \varepsilon_\phi)$
and the decay width is then calculated by
$$
\Gamma (\phi \to S \gamma) ~=~
      {{\alpha q} \over {3 m_\phi^2}}
   ~   | \tilde M |^2
{}~~~~,~~~~
\tilde M  = \tilde M_a +
              \tilde M_b +
              \tilde M_c +
              \tilde M_d
{}~~~.
\eqno{(9)}
$$

Our numerical results for $\Gamma(\phi\rightarrow f_0\gamma)$
are shown in Fig. 2 as a function of the $K\bar K$-molecule
radius. For the numerical evaluations, the coupling
constants $g_\phi=4.57$ and $g^2/4\pi=0.6$ GeV$^2$ [11]
are used.
In the pointlike limit ($R_{K\bar K}\rightarrow 0$),
we obtain the width $6.3\times 10^{-4}$ MeV,
which exactly agrees with
the value (marked by $\times$ in Fig. 2)
obtained by using the analytical expressions
in Ref. 11.
However, if the $K\bar K$-molecule size
1.2 fm (the Toronto $K\bar K$ molecule) is used,
we obtain the width $1.8\times 10^{-4}$ MeV,
which is significantly
smaller than the point-like result.
Using the total $\phi$ width 4.41 MeV, we
get the branching ratio
$B.R.(\phi\rightarrow f_0\gamma)=4 \times 10^{-5}$.
In the $K\bar K$-type picture for $a_0$, the decay width
for $\phi\rightarrow K\bar K\gamma \rightarrow a_0 \gamma$
is roughly the same as the one for
$\phi\rightarrow f_0\gamma$.

\vspace{8.5cm}

Fig. 1 $\phi\rightarrow K\bar K\gamma \rightarrow S\gamma$
       processes.
\hspace{1.6cm}
Fig. 2 $\Gamma(\phi\rightarrow f_0\gamma)$ versus $R_{K\bar K}$.

\vfill\eject

\noindent
{4. \underline{Comments on a CP background problem
                    and on the OZI rule}}

\vspace{0.2cm}

Because the photon charge conjugation
is negative, $K_S K_S$ or $K_L K_L$
are produced in the $\phi$ radiative decays
$\phi\rightarrow S\gamma
     \rightarrow K^0 \bar K^0\gamma$.
The ratio of CP violation parameters
$\varepsilon '/\varepsilon$ will be measured
by using the decay $\phi \rightarrow K_S K_L$.
If the branching ratios for the radiative
decays are very large, it becomes a serious
problem for measuring $\varepsilon '/\varepsilon$
unless we find a method of excluding
the radiative processes.
Prior to our paper [1], there were a number of
publications on this topic. However, it is rather
surprizing to find that the branching ratio varies
from $10^{-5}$ to $10^{-9}$. This large fluctuation
is due in part to errors and in part to differences
in modelling [11].
(See the Brown-Close preprint in Ref. 11 for discussions
 on the previous calculations.)
The resonant contribution to
$\phi\rightarrow K^0\bar K^0\gamma$ is calculated by
using the results in the previous sections and
also the Breit-Wigner form for the
$S$ propagation:
$$
 {{d\Gamma (\phi\rightarrow S\gamma
                \rightarrow K^0 \bar K^0 \gamma)} \over
  {dp_s^2}}
={ 1 \over {(4\pi)^2}} ~
\sqrt{1- {{4m_{_{K^0}}^2} \over {p_s^2}}} ~
  {{g^2} \over {(p_s^2-m_s^2)^2 +m_s^2 \Gamma_s^2}} ~
    \Gamma(\phi\rightarrow S\gamma) ~~~,
\eqno{(10)}
$$
where $p_s$ is the scalar-meson $S$ momentum.
We should note that there is a destructive
interference between the $f_0$ and $a_0$ amplitudes.
{}From the above equation and the results in sections 2 and 3,
we obtain
$B.R.(\phi\rightarrow K^0\bar K^0\gamma)\lapproxeq 10^{-7}$
[1,11]. This is too small to be a serious background to
the CP-violation experiment.

Our investigations on the radiative decays
suggest an interesting point on the OZI
rule.
If $f_0=(u\bar u+d\bar d)/\sqrt{2}$, the decay
$\phi\rightarrow f_0\gamma$ should vanish
in the lowest order and
the (OZI-violating) $K\bar K$-loop contribution
provides a small correction.
If $f_0=s\bar s$, we obtained the branching ratio
for the ``direct'' process:
$B.R.[\phi\rightarrow f_0(s\bar s)\gamma]\approx 10^{-5}$.
According to the OZI rule,
this is supposed to be much larger than
the OZI-violating decay
$\phi\rightarrow K\bar K\gamma \rightarrow f_0\gamma$.
If the $K\bar K$ system is diffuse ($R_{K\bar K}>$2 fm),
the loop process gives $B.R.(\phi\rightarrow f_0\gamma)<10^{-5}$
from Fig. 2. In this case, the empirical OZI rule is valid.
This is due to the poor spatial overlap between
the $K\bar K$ system and the $\phi$.
Because the point-like calculation fails to
take into account this confinement scale,
the obtained rate is $10^{-4}$ and the OZI rule
is invalid.

Next, assuming the $s\bar s$ for $f_0$, we consider
the decay connected by a $q\bar q s\bar s$
intermediate state.
There are two types of contributions
from the $q\bar q s\bar s$ loops
at the quark level.
First, there are the diffuse
contributions which can arise from hadronic
loops corresponding to nearby thresholds,
in this case from $K\bar K$.
Then, there are short distance contributions.
A realistic calculation of such contributions
should include a large set of hadronic loops, and
it was found that such hadronic loop contributions
tend to cancel each other [15]. Therefore,
the incompleteness of the cancellation of OZI-violating
hadronic loops is due
to nearby thresholds.

\vfill\eject

\noindent
{5. \underline{Conclusions}}

\vspace{0.2cm}

We investigated the $\phi$ radiative decays
into the scalar mesons $f_0$(975) and $a_0$(980).
We found that the decay widths vary widely:
$B.R.=10^{-4}-10^{-6}$
depending on these meson substructures. Therefore,
it should be possible to discriminate among various
models ($q\bar q$, $qq\bar q\bar q$, $K\bar K$, glueball)
by measuring these decay widths at future
$\phi$ factories.
However, our naive investigation is merely
a starting point. Much detailed analyses are needed
for the various substructure effects on
the radiative decays.

We found that the radiative decay
$\phi\rightarrow K^0\bar K^0\gamma$
is not a serious background to
the CP-violation studies at the $\phi$ factories
because the obtained branching ratio
$B.R.(\phi\rightarrow K^0\bar K^0\gamma) \lapproxeq 10^{-7}$
is small.

\vspace{0.7cm}

\begin{center}
{\underline{Acknowledgment}} \\
\end{center}

This research was partly supported by
the Deutsche Forschungsgemeinschaft (SFB 201).
S.K. thanks F. E. Close and N. Isgur for
discussions on the $\phi$ radiative decays.
S.K. thanks people at the Institute for
Nuclear Study in Tokyo
for their financial support
for his staying at the INS.

$~~~$

\noindent
* present address. E-mail: KUMANO@VKPMZP.KPH.UNI$-$MAINZ.DE

\vspace{0.7cm}

\begin{center}
{\underline{References}} \\
\end{center}

\vspace{-0.30cm}
\vspace{-0.38cm}
\begin{description}{\leftmargin 0.0cm}
\item{[1]}
F. E. Close, N. Isgur, and S. Kumano,
preprint (RAL-92-026, CEBAF-TH-92-13, IU/NTC-92-16),
to be published in Nucl. Phys. B.

\vspace{-0.38cm}
\item{[2]}
S. Godfrey and N. Isgur, Phys. Rev. {\bf D32}, 189 (1985).

\vspace{-0.38cm}
\item{[3]}
R. Kokoski and N. Isgur, Phys. Rev. {\bf D35}, 907 (1987);
S. Kumano and V. R. Pandharipande, Phys. Rev. {\bf D38}, 146 (1988)
and references therein.

\vspace{-0.38cm}
\item{[4]}
The Particle Data Group, Phys. Lett. {\bf 239B}, 1 (1990).
In our numerical analysis, data in the 1990-version
are used. The most recent one is in
Phys. Rev. {D45}, I.1 (1992), where
$\Gamma(f_0)=47$ MeV is listed.

\vspace{-0.38cm}
\item{[5]}
For example, see
B. D. Serot and J. D. Walecka,
         Adv. Nucl. Phys. {\bf 16}, 1 (1986).
For the ``$\sigma$'' width, see Refs. 6 and 3.

\vspace{-0.38cm}
\item{[6]}
J. Weinstein and N. Isgur,
        Phys. Rev. Lett. {\bf 48}, 659 (1982);
        Phys. Rev. {\bf D27}, 588 (1983);
                   {\bf D41}, 2236 (1990);
K. L. Au, D. Morgan, and M. R. Pennington,
        Phys. Lett. {\bf 167B}, 229 (1986);
                    {\bf 258B}, 444 (1991);
D. Lohse, J. W. Durse, K. Holinde, and J. Speth,
        Phys. Lett. {\bf 234B}, 235 (1990);
        Nucl. Phys. {\bf A516}, 513 (1990);
V. Mull, J. Wambach, and J. Speth,
        Phys. Lett. {\bf 286B}, 13 (1992).

\vspace{-0.38cm}
\item{[7]}
T. Barnes, Phys. Lett. {\bf 165B}, 434 (1985).

\vspace{-0.38cm}
\item{[8]}
Z. P. Li, F. E. Close, and T. Barnes,
           Phys. Rev. {\bf D43}, 2161 (1991);
Z. P. Li and F. E. Close, Z. Phys. {\bf C 54}, 147 (1992).
T. N. Truong, talk at the Hadron '89 Workshop,
              Ajaccio, France, Sept. 23-27, 1989.

\vspace{-0.38cm}
\item{[9]}
M. Piccolo, talk given at this conference;
D. B. Cline, Comm. Part. Nucl. Phys. {\bf 20}, 241 (1992);
M. Fukawa et al., KEK report 90-12 (1990);
L. M. Barkov et al., in Proc. of the 1991 IEEE
                        Part. Acc. Conf.,
                        San Francisco (1991).

\vspace{-0.38cm}
\item{[10]}
I. Dunietz, J. Hauser, and J. L. Rosner,
    Phys. Rev. {\bf D35}, 2166 (1987);
J. Bernab\'eu, F. J. Botella, and J. Rold\'an,
    Phys. Lett. {\bf 211B}, 226 (1988).

\vspace{-0.38cm}
\item{[11]}
S. Nussinov and T. N. Truong,
    Phys. Rev. Lett. {\bf 63}, 1349 \& 2003 (1989);
N. N. Achasov and V. N. Ivanchenko,
    Nucl. Phys. {\bf B315}, 465 (1989);
N. Paver and Riazuddin,
    Phys. Lett. {\bf 246B}, 240 (1990);
J. L. Lucio M. and J. Pestieau,
    Phys. Rev. {\bf D42}, 3253 (1990); {\bf D43}, 2447 (1991);
Comparisons and comments on these results are given in
N. Brown and F. E. Close, preprint RAL-91-085 and Ref. 1.

\vspace{-0.38cm}
\item{[12]}
After completion of our research, we find
other recent investigations on the $\phi$
radiative decays:
P. J. Franzini, W. Kim, and J. Lee-Franzini,
    Phys. Lett. {\bf 287B}, 259 (1992);
A. Bramon, G. Colangelo, and M. Greco,
    Phys. Lett. {\bf 287B}, 263 (1992).

\vspace{-0.38cm}
\item{[13]}
R. L. Jaffe, Phys. Rev. {\bf D15}, 267 \& 281 (1977);
                        {\bf D17}, 1444 (1978);
R. L. Jaffe and K. Johnson, Phys. Lett. {\bf 60B}, 201 (1976);
R. L. Jaffe and F. E. Low, Phys. Rev. {\bf D19}, 2105 (1979);
N. N. Achasov, S. A. Devyanin, and G. N. Shestakov,
        Phys. Lett. {\bf 96B}, 168 (1980);
        Sov. J. Nucl. Phys. {\bf 32}, 566 (1980);
        Sov. Phys. Usp. {\bf 27}, 161 (1984);
N. A. T\"ornqvist, Phys. Rev. Lett. {\bf 49}, 624 (1982).

\vspace{-0.38cm}
\item{[14]}
L. Heller, S. Kumano, J. C. Martinez, and E. J. Moniz,
           Phys. Rev. {\bf C35}, 718 (1987).

\vspace{-0.38cm}
\item{[15]}
P. Geiger and N. Isgur,
           Phys. Rev. Lett. {\bf 67}, 1066 (1991);
           Phys. Rev. {\bf D44}, 799 (1991).

\end{description}
\end{document}